\documentclass[11pt,a4paper,useAMS,usenatbib]{emulateapj}
\usepackage{epsfig}
\usepackage{amsmath}
\usepackage{natbib}
\usepackage{ulem} 
\bibpunct{(}{)}{;}{a}{}{,}

\usepackage[dvipsnames]{xcolor}

\begin{document}

\title{Ultra-diffuse galaxies in the Coma cluster: \\ Probing their origin and AGN occupation fraction}

\author{Orsolya E. Kov\'acs\altaffilmark{1,2,3}, \'Akos
Bogd\'an\altaffilmark{1}, Rebecca E. A. Canning\altaffilmark{4}}
\affil{\altaffilmark{1}Harvard Smithsonian Center for Astrophysics, 60
Garden Street, Cambridge, MA 02138, USA; orsolya.kovacs@cfa.harvard.edu}
\affil{\altaffilmark{2}Konkoly Observatory, MTA CSFK, H-1121 Budapest,
Konkoly Thege M. {\'u}t 15-17, Hungary}
\affil{\altaffilmark{3}E{\"o}tv{\"o}s University, Department of
Astronomy, Pf. 32, 1518, Budapest, Hungary} \affil{\altaffilmark{4}Kavli
Institute for Particle Astrophysics and Cosmology, Stanford University,
452 Lomita Mall, Stanford, CA 94305-4085, USA}
\shorttitle{ULTRA-DIFFUSE GALAXIES IN THE COMA CLUSTER}
\shortauthors{KOV\'ACS ET AL.}

\begin{abstract}
Ultra-diffuse galaxies (UDGs) exhibit low surface brightness, but their optical extent is comparable to Milky Way-type galaxies. In this work, we utilize \textit{Chandra} X-ray observations of 404 UDGs in the Coma cluster and address two crucial goals. First, we constrain the  formation scenario of UDGs by probing the X-ray emission originating from diffuse gas and from the population of unresolved low-mass X-ray binaries (LMXBs) residing in globular clusters (GCs). It is expected that both the luminosity of the hot gas and the number of GC and hence the luminosity from GC-LMXBs are proportional to the total dark matter halo mass. We do not detect statistically significant emission from the hot gas or from GC-LMXBs. The upper limits on the X-ray luminosities suggest that the bulk of the UDGs reside in low-mass dark matter halos, implying that they are genuine dwarf galaxies. This conclusion agrees with our previous results obtained for isolated UDGs, arguing that UDGs are a homogenous population of galaxies. Second, we constrain the active galactic nuclei (AGN) occupation fraction of UDGs for the first time. To this end, we cross-correlate the position of detected X-ray sources in the Coma cluster with the position of UDGs. We identify two UDGs that have a luminous X-ray source at $3.0\arcsec$ and $3.2\arcsec$ from the center of the galaxies, which could be off-center AGN. However, Monte Carlo simulations suggest that one of these sources could be the result of spatial coincidence with a background AGN. Therefore, we place an upper limit of $\lesssim0.5\%$  on the AGN occupation fraction of UDGs.  \\ 
\end{abstract}



\section{Introduction}
\label{sec:intro}

\begin{figure*}[ht!]
	\begin{center}
		\leavevmode
		\epsfxsize=0.48\textwidth \epsfbox{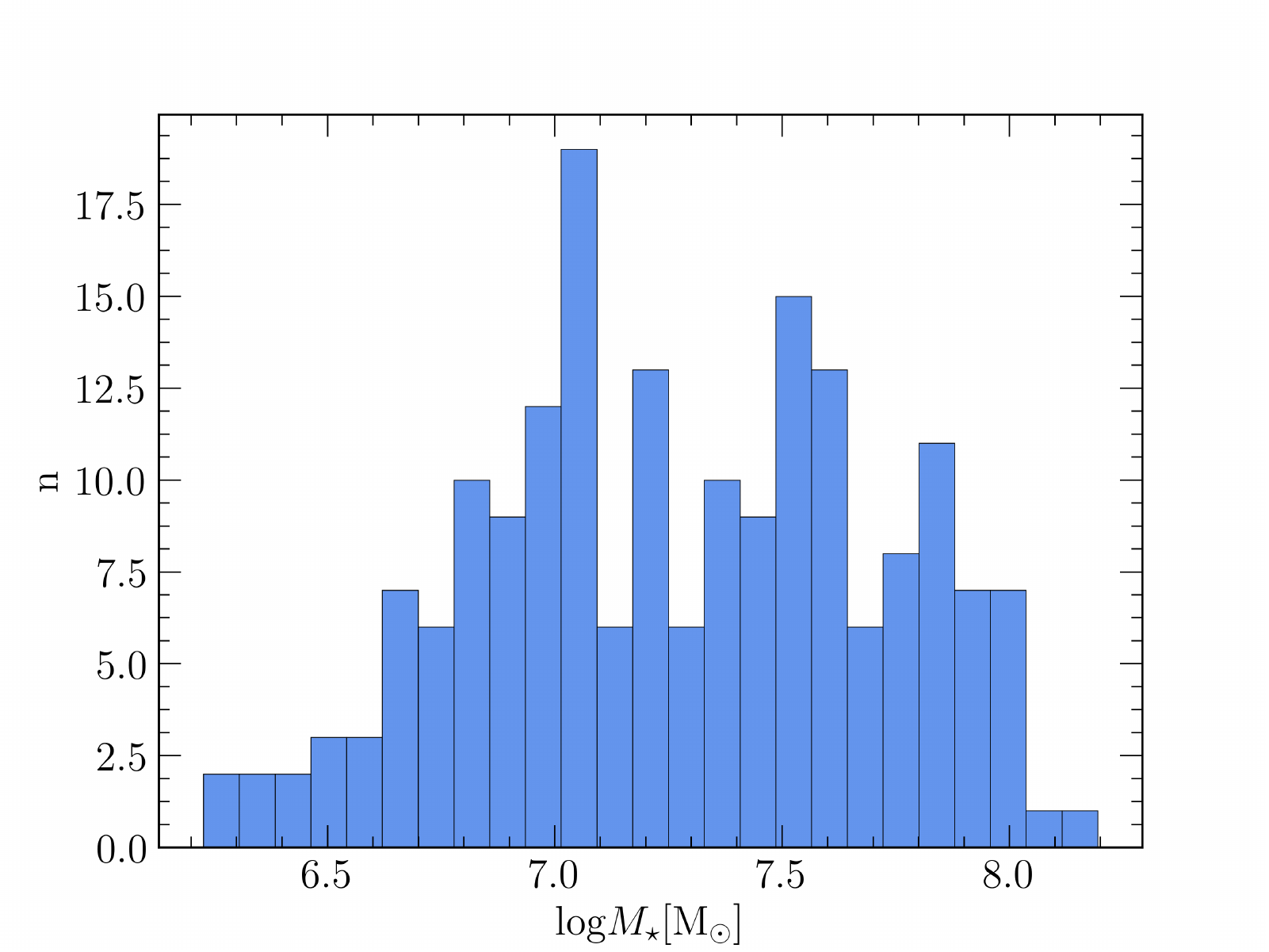}
		\epsfxsize=0.48\textwidth \epsfbox{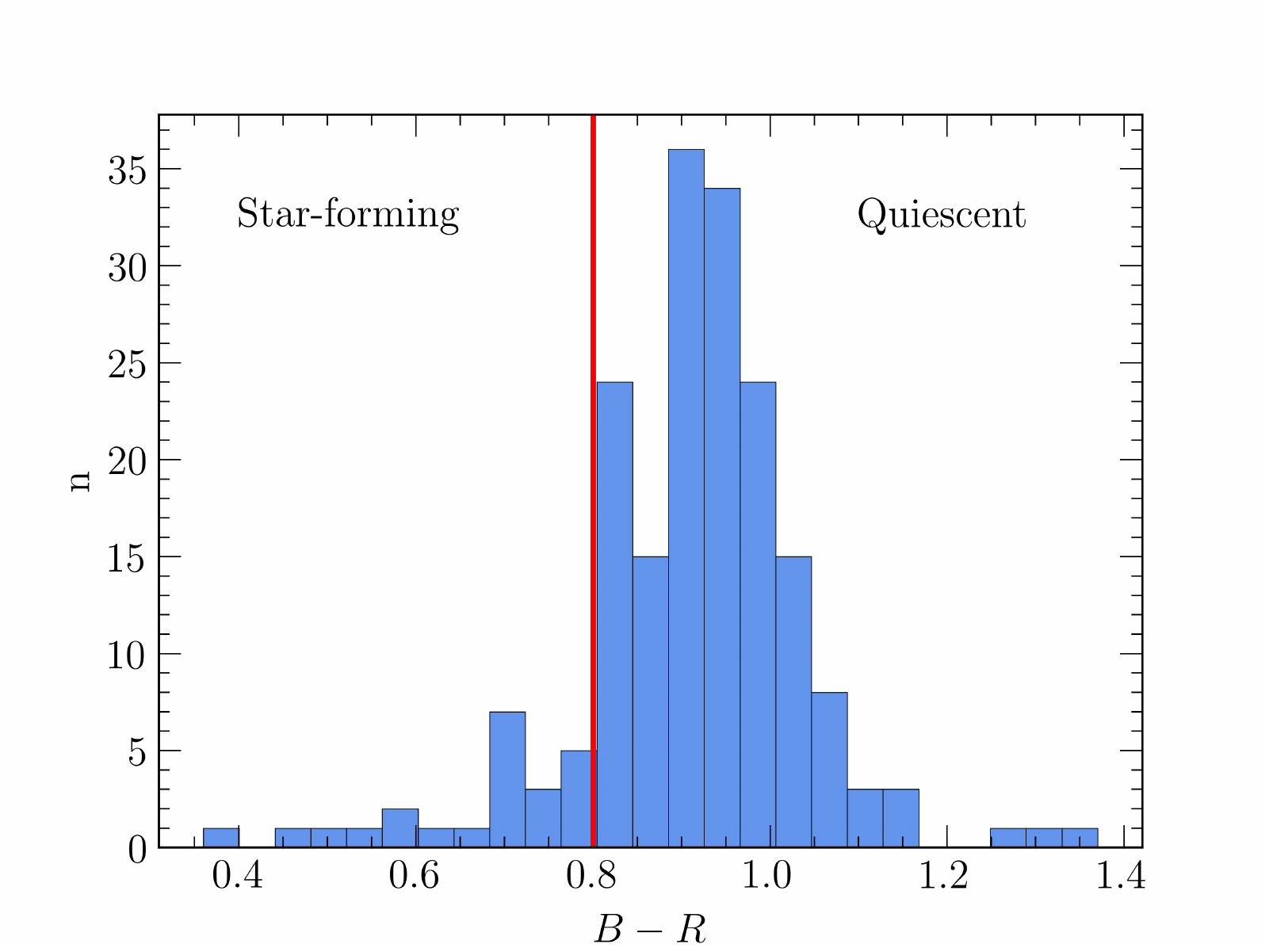}
		\vspace{0cm}
		\caption{{\it Left:} The stellar mass ($M_{\star}$) distribution of the analyzed UDGs calculated using the $R$ band magnitudes and $B-R$ color indices from \citet{2016ApJS..225...11Y}. For the computation, we assume that the galaxies are at the distance of the Coma cluster. {\it Right:} The $B-R$ color index histogram of the analyzed UDGs. The distribution of our sample is consistent with the full list of UDGs \citep{2016ApJS..225...11Y} with UDGs being mainly on the red sequence. This suggests the quiescent nature of the majority of the  UDGs. The red vertical line denotes the boundary between quiescent and star-forming color regime \citep{2015ApJ...807L...2K}.} 
		\label{fig:1}
	\end{center}
\end{figure*}

Ultra-diffuse galaxies (UDGs) are a curious population of galaxies that have extremely low central surface brightness ($\mu_{0}(g)\gtrsim24 \ \rm{mag \ arcsec^{-2}}$), but large  effective radii  ($r_{\rm{eff}} \gtrsim 1.5  \ \rm{kpc}$). Although these galaxies were known for decades as  Low Surface Brightness Galaxies or Low-Mass Cluster Galaxies \citep[e.g.][]{1984AJ.....89..919S,1988ApJ...330..634I,2003AJ....125...66C}, recent observational studies re-discovered (and re-named) these galaxies. Most notably, the Dragonfly Telephoto Array \citep{2014PASP..126...55A}, which was specifically designed to resolve structures at low surface brightness levels, detected a large population of UDGs in a galaxy cluster \citep{2015ApJ...798L..45V}. This discovery was followed by searches of UDGs in other environments \citep[e.g.][]{2015ApJ...809L..21M,2016AJ....151...96M,2016ApJ...833..168M,2017MNRAS.468.4039R,2017MNRAS.467.3751B,2017ApJ...842..133L,2018ApJ...857..104G}. These studies revealed that UDGs are common in galaxy clusters, galaxy groups, and as field galaxies 

The Coma cluster hosts a large UDG population. While initial observations with the Dragonfly Telephoto Array discovered 47 UDGs  \citep{2015ApJ...798L..45V}, in a follow-up study, carried out by
surveying deep Suprime-Cam/Subaru R-band images of Subaru data archive,  \cite{2016ApJS..225...11Y} detected a population of 854 UDG candidates in the Coma cluster. Given the massive nature and the relative proximity of the Coma cluster, the galaxy cluster is well-studied and has rich multi-wavelength observations. Specifically, it was the target of multiple \textit{Chandra} campaigns, which observations explored the galaxy cluster out to $\sim1\degr$ radius. These observations offer a unique opportunity to study UDGs residing in the Coma cluster. In this study, we employ \textit{Chandra} observations of Coma cluster UDGs to address two critical science goals: (1) we probe the formation scenarios and (2) measure the AGN occupation fraction of UDGs.

The curious properties of UDGs can be explained by different formation scenarios. UDGs may be the descendants of massive galaxies, which lost their gas content at high redshift \citep{2015ApJ...798L..45V,2016ApJ...828L...6V}. This, in turn, halted the star formation, resulting in low-surface brightness galaxies residing in massive dark matter halos. Alternatively, UDGs could be genuine dwarf galaxies\citep{2020A&A...633L...3C}, whose large spatial extent is due to feedback-driven gas outflows \citep[e.g.][]{2016ApJ...830...23B,2016MNRAS.459L..51A}. In this scenario, most physical characteristics of UDGs, such as their dark matter halo mass, is similar to dwarf galaxies. 
 
Measuring the dark matter halo mass of UDGs offers a robust method to distinguish between potential formation scenarios. If UDGs originate from massive galaxies, they will reside in massive dark matter halos with virial radius of $M_{\mathrm{vir}} \gtrsim 10^{11} \ \rm{M_{\odot}} $. If, however, UDGs are puffed-up dwarf galaxies, they are expected to live in low-mass dark matter halos  ($M_{\mathrm{vir}} \lesssim 3\times 10^{10} \ \rm{M_{\odot}} $). X-ray observations provide a powerful tool to measure the total gravitating mass of galaxies. Indeed, empirical correlations revealed that the luminosity of the gaseous X-ray halo around galaxies is proportional to the galaxy's total gravitating mass \citep{2013ApJ...776..116K,2018ApJ...857...32B}. Additionally, comprehensive X-ray studies demonstrated that low-mass galaxies, such as dwarf galaxies, cannot retain a significant amount of X-ray emitting gas due to their shallow potential well \citep{2006ApJ...653..207D}. By utilizing the X-ray luminosity--total gravitating mass relation, we constrained the formation scenarios of isolated UDGs  \citep{2019ApJ...879L..12K}. Specifically, we analyzed \textit{XMM-Newton} X-ray observations to probe the X-ray characteristics of a sample of isolated UDG candidates
identified in the Subaru data.
We did not detect statistically significant X-ray emission from the individual galaxies or from the stacked set of galaxies. The absence of the detection demonstrates the lack of an X-ray halo, thereby suggesting that most isolated UDGs are puffed-up dwarf galaxies. 

To follow-up on our previous study, we  extend our analysis to UDGs residing in the Coma cluster \citep{2016ApJS..225...11Y}. We carry out a similar analysis to that presented in \citet{2019ApJ...879L..12K}. Specifically, we study the gaseous X-ray halo of Coma cluster UDGs to constrain whether their luminosity is consistent with a high-mass or a low-mass dark matter halo. As a complementary approach, we also probe the dark matter halo mass of UDGs through their globular cluster (GC) population. Several UDGs were demonstrated to host a large number of GCs \citep{2017ApJ...844L..11V}, suggesting that these UDGs reside in massive dark matter halos  \citep{2020AJ....159...56B}. GCs are known to effectively form low-mass X-ray binaries (LMXBs) through dynamical processes \citep{2007MNRAS.380.1685V}. These LMXBs are copious sources of X-ray emission. Hence, by probing the X-ray luminosity associated with GC-LMXBs, we can probe whether the bulk of UDGs hosts a large GC population, as would be expected if they reside in a massive dark matter halo. 

The exquisite \textit{Chandra} X-ray observations of the Coma cluster provide the opportunity to probe whether UDGs host AGN. Therefore, for the first time, we constrain the AGN occupation fraction of UDGs. X-ray observations offer a powerful method to search for AGN as they are sensitive to weakly accreting black holes. In the past decade, several studies explored the AGN occupation fraction of dwarf galaxies \citep[e.g.][]{2013ApJ...775..116R,2015ApJ...799...98M}. Both X-ray studies and optical line diagnostics investigations revealed that a small fraction of dwarf galaxies host AGN, whose mass is in the range of $10^4-10^5 \ \rm{M_{\odot}}$. While most previous studies focused on dwarf galaxies, the population of UDGs was not considered. Given that the mass of BHs is interconnected with the stellar mass and dark matter halo mass of the host galaxy \citep[e.g.][]{2013ARA&A..51..511K,2015ApJ...800..124B}, it is interesting to probe whether the AGN occupation fraction of UDGs is similar to dwarf or more massive galaxies. 

For the distance of Coma cluster, we assumed $D=103$~Mpc. At this distance $1\arcsec$ corresponds to $0.476$~kpc. The Galactic absorption toward Coma cluster is $9.3\times10^{19} \ \rm{cm^{-2}}$ \citep{2010atnf.prop.3373B}. In this paper, we used standard $\Lambda$-CDM cosmology with   $H_0 = 70  \ \rm{km \ s^{-1} \ Mpc^{-1}},  \Omega_M=0.3$, and $\Omega_{\Lambda}=0.7$. The errors quoted in the paper are $1\sigma$ uncertainties unless otherwise noted.

This paper is structured as follows. Section 2 describes the sample of UDGs detected in the Subaru survey.
In Section 3, we introduce the analyzed data and discuss its reduction. In Section 4, we place constraints on the formation scenarios of UDGs. Our result  on the AGN occupation fraction is presented in Section 5. We discuss the results in Section 6 and summarize in Section 6. 

\begin{table*}[!t]
\begin{center}
\caption{List of analyzed \textit{Chandra} observations}
\begin{minipage}{18cm}
\renewcommand{\arraystretch}{1.3}
\centering
\begin{tabular}{ c|c|c|c||c|c|c|c}
\hline
Obs.\,ID & $t_{\rm exp} [ks]$ & Detector & Obs.\,Date & Obs.\,ID & $t_{\rm exp} [ks]$ & Detector & Obs.\,Date \\
(1) & (2) & (3) & (4) & (1) & (2) & (3) & (4)\\
\hline \hline
555	&	8.67	&	ACIS-01237	&	1999-11-04	&	18276	&	84.19	&	ACIS-0123	&	2016-03-05	\\
556	&	9.65	&	ACIS-23678	&	1999-11-04	&	18761	&	47.45	&	ACIS-0123	&	2016-03-13	\\
1086	&	9.64	&	ACIS-23678	&	1999-11-04	&	18791	&	34.62	&	ACIS-0123	&	2016-03-08	\\
1112	&	9.65	&	ACIS-01237	&	1999-11-04	&	18792	&	21.25	&	ACIS-0123	&	2016-03-09	\\
1113	&	9.65	&	ACIS-01237	&	1999-11-04	&	18793	&	76.95	&	ACIS-0123	&	2016-03-10	\\
1114	&	9.05	&	ACIS-01237	&	1999-11-04	&	18794	&	29.30	&	ACIS-0123	&	2016-03-14	\\
2941	&	62.91	&	ACIS-0123	&	2002-04-11	&	18795	&	27.78	&	ACIS-0123	&	2016-03-17	\\
4724	&	59.68	&	ACIS-012367	&	2004-02-17	&	18796	&	39.56	&	ACIS-0123	&	2016-03-18	\\
8188	&	28.66	&	ACIS-23567	&	2007-03-14	&	18797	&	54.36	&	ACIS-0123	&	2016-03-19	\\
9714	&	29.65	&	ACIS-01236	&	2008-03-20	&	18798	&	12.91	&	ACIS-0123	&	2016-03-20	\\
10672	&	28.54	&	ACIS-23567	&	2009-03-15	&	19909	&	9.93	&	ACIS-3678	&	2016-11-01	\\
10921	&	4.99	&	ACIS-235678	&	2009-06-27	&	19910	&	9.92	&	ACIS-23678	&	2016-11-02	\\
12887	&	43.44	&	ACIS-01236	&	2010-11-11	&	19911	&	9.91	&	ACIS-23678	&	2016-11-03	\\
13993	&	39.56	&	ACIS-01236	&	2012-03-21	&	19912	&	9.92	&	ACIS-23678	&	2016-11-07	\\
13994	&	81.99	&	ACIS-01236	&	2012-03-19	&	19998	&	31.66	&	ACIS-0123	&	2017-03-13	\\
13995	&	62.99	&	ACIS-01236	&	2012-03-14	&	20010	&	58.30	&	ACIS-0123	&	2017-02-18	\\
13996	&	123.06	&	ACIS-01236	&	2012-03-27	&	20011	&	44.49	&	ACIS-0123	&	2017-02-19	\\
14406	&	24.76	&	ACIS-01236	&	2012-03-15	&	20027	&	20.81	&	ACIS-0123	&	2017-03-12	\\
14410	&	78.54	&	ACIS-01236	&	2012-03-22	&	20028	&	43.51	&	ACIS-0123	&	2017-04-13	\\
14411	&	33.64	&	ACIS-01236	&	2012-03-20	&	20029	&	21.79	&	ACIS-0123	&	2017-03-08	\\
14415	&	34.53	&	ACIS-012367	&	2012-04-13	&	20030	&	33.15	&	ACIS-0123	&	2017-04-11	\\
18234	&	19.79	&	ACIS-23678	&	2017-03-29	&	20031	&	10.93	&	ACIS-0123	&	2017-03-11	\\
18235	&	29.67	&	ACIS-3678	&	2017-03-21	&	20037	&	16.86	&	ACIS-0123	&	2017-03-16	\\
18236	&	9.94	&	ACIS-3678	&	2016-11-01	&	20038	&	49.92	&	ACIS-0123	&	2017-03-18	\\
18237	&	9.93	&	ACIS-23678	&	2016-11-03	&	20039	&	21.89	&	ACIS-0123	&	2017-03-19	\\
18271	&	54.36	&	ACIS-0123	&	2017-03-15	&	20049	&	19.79	&	ACIS-23678	&	2017-03-29	\\
18272	&	19.82	&	ACIS-0123	&	2016-03-08	&	20050	&	13.87	&	ACIS-23678	&	2017-03-29	\\
18273	&	28.32	&	ACIS-0123	&	2017-02-15	&	20051	&	14.86	&	ACIS-23678	&	2017-03-31	\\
18274	&	46.47	&	ACIS-0123	&	2017-03-06	&	20052	&	23.74	&	ACIS-23678	&	2017-04-16	\\
18275	&	49.43	&	ACIS-0123	&	2016-03-16	&&&&	\\																			
															
 \hline
\end{tabular} 
\end{minipage}
\end{center}
Columns are as follows: (1) \textit{Chandra} ID of the observation; (2) Exposure time; (3) Detector name; (4) Date of observation.
\label{tab:1}
\end{table*}

\section{Ultra-diffuse galaxies in the Coma cluster}
\label{sec:sample}

To study a large and homogeneous sample of UDGs, we rely on UDGs residing in the Coma cluster. Coma was the first galaxy cluster, in which the Dragonfly Telephoto Array discovered a significant population of UDGs \citep{2015ApJ...798L..45V}. The initial Dragonfly survey was followed by a deeper study \citep{2015ApJ...804L..26V}
and the Subaru Suprime-Cam survey,
which revealed 854 UDG candidates in the Coma cluster \citep{2016ApJS..225...11Y}.
We note that the bulk of these galaxies should be considered as UDG candidates, as spectroscopic distance measurements are only available for a fraction of them. Due to the faint nature of UDGs and the large angular extent of the Coma cluster, it is demanding to measure the radial velocities of a large set of UDGs. However, using optical spectroscopic data from 10-m class telescopes, the radial velocity of small samples of UDG candidates were measured. These studies confirmed that most UDG candidates are members of the Coma cluster, and only a few of them reside behind the galaxy cluster \citep{2016ApJ...828L...6V,2017ApJ...844L..11V,2017ApJ...838L..21K,2018MNRAS.479.3308A,2018ApJ...859...37G}. While these measurements are encouraging, some of the candidate UDGs without accurate radial velocity measurements may not reside in the Coma cluster. This, in turn, could imply that their true effective radius is less than $1.5$~kpc if they are projected to the Coma cluster and the distance of the galaxies is significantly less than $D<103$~Mpc. 

In this work, we study the publicly available UDG candidates identified by \citet{2016ApJS..225...11Y}. These galaxies are distributed in a 1$\degr$ region around the center of the Coma cluster. By utilizing the archival  \textit{Chandra} observations in the footprint of the Coma cluster and excising the central $4\arcmin$ region (Section \ref{sec:data}), we find that 404 UDGs candidates have \textit{Chandra} observations.
 
The Subaru survey provides magnitudes in $B$ and $R$ bands. To derive the stellar mass of the galaxy sample, we compute the $R$-band mass-to-light ratios using the $B-R$ color indices and rely on the results of \citet{2003ApJS..149..289B}. Most UDGs have stellar masses in the  range of $10^6 - 10^8 \ \rm{M_{\odot}}$. The stellar mass distribution of the galaxies is presented in the left panel of Figure \ref{fig:1}.  Based on the $B-R$ color indices we establish that most UDGs in our sample are red galaxies (right panel of Figure \ref{fig:1}) and have little to no star-formation. This conclusion is in good agreement with that obtained for the full sample  \citep{2016ApJS..225...11Y}.

\begin{figure*}[ht!]
	\begin{center}
		\leavevmode
		\epsfxsize=0.96\textwidth \epsfbox{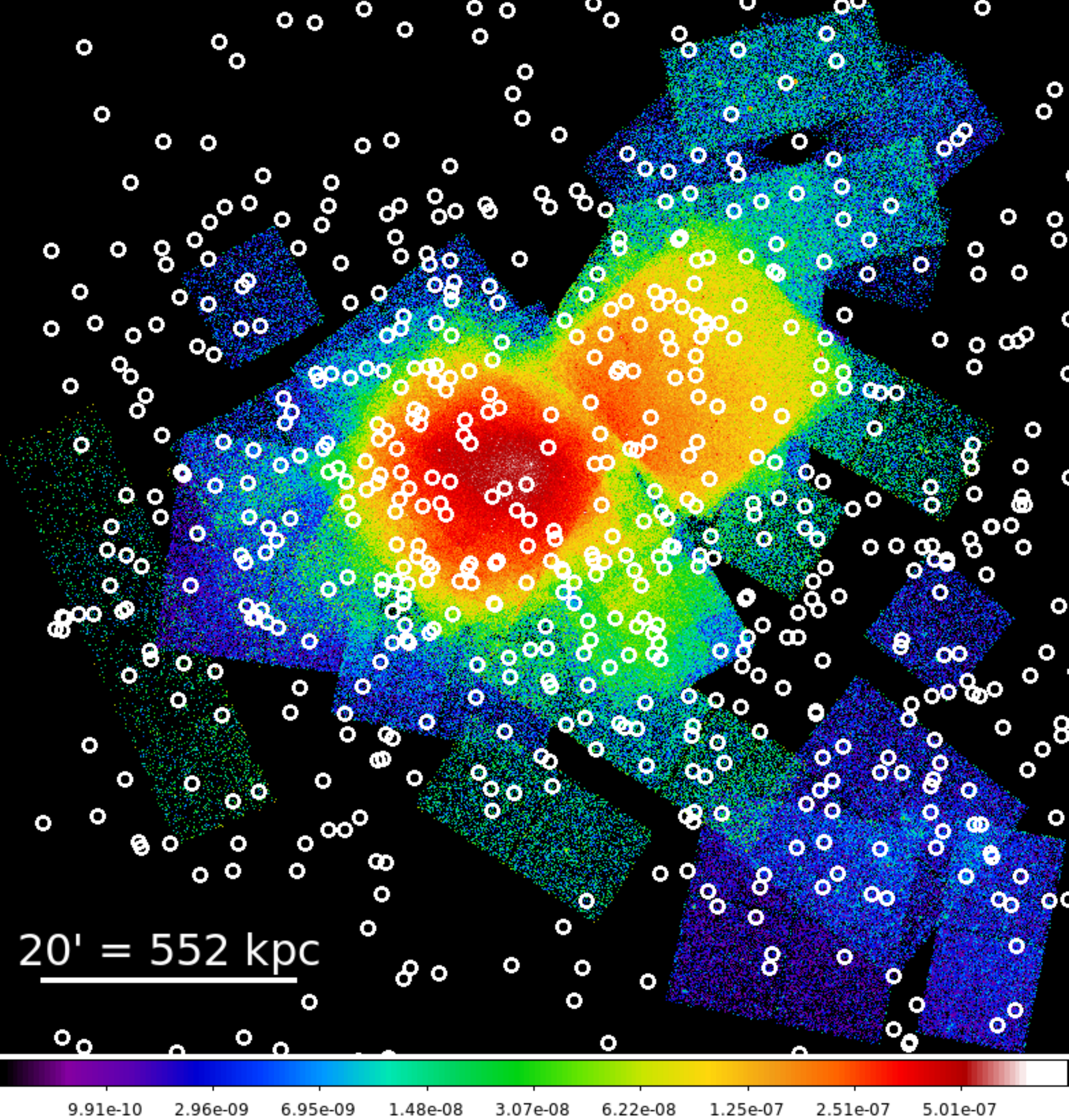}
		\caption{Merged $0.5-7$~keV band \textit{Chandra} X-ray image image of the Coma cluster. The mosaic includes 59 \textit{Chandra} ACIS imaging observations within a $1\degr$ radius of the center of the cluster. The individual images were combined and exposure correction is applied. The image is smoothed with a Gaussian with a kernel size of 4 pixels. The small white circles show the position of the UDGs identified by the Subaru survey of the Coma cluster \citep{2016ApJS..225...11Y}. }
	\label{fig:5}
	\end{center}
\end{figure*}

\section{Data analysis}
\label{sec:data}

In this work, we focus on \textit{Chandra} ACIS imaging observations of the Coma cluster. To identify suitable data, we searched for publicly available observations within a $1\degr$ (or $1.71$ Mpc) radius of the cluster center. We identified 59 observations with a total exposure time of $1954.4$\,ks. These observations include 17 ACIS-S and 42 ACIS-I observations. The analyzed set of \textit{Chandra} observations along with their exposure times are listed in Table~\ref{tab:1}. 

We prepared and analyzed the data using standard \textsc{CIAO} tools (version 4.11) and used the most recent Calibration Database (CalDB version 4.8.4.1). To apply the latest calibration files on the data, we reprocessed all observations using the \texttt{chandra\_repro} tool. Given that the main focus of the paper is to identify luminous X-ray sources, we did not filter flare contaminated time intervals, as the longer exposure and the resulting higher sensitivity outweigh the effect of a potentially higher background level. 

From the event lists, we constructed images in the $0.5-1.2$~keV (soft) and $0.5-7$~keV (broad) bands to study the emission from hot gas and LMXBs, respectively. Given the characteristic X-ray spectrum of the gaseous emission and the population of LMXBs, these energy ranges are ideal to maximize the signal-to-noise ratios. We generated exposure maps for each observation assuming a different spectrum for each band, which corrects for vignetting effects and allows us to convert the counts to flux. For the soft band images, we assume an optically-thin thermal plasma  emission model (APEC model in XSpec) with a temperature of $0.2$~keV and a metallicity of $Z = 0.2 \ \rm{Z_{\odot}}$, which typically describes the gaseous emission of low-mass elliptical and spiral galaxies \citep[e.g.][]{2011MNRAS.418.1901B,2016ApJ...826..167G}. Although individual galaxies exhibit some variations in their best-fit gas temperature and metallicity, these potential variations do not affect our results, especially when investigating large samples of galaxies.  For the broad-band images, we assume a power-law model with a slope of $\Gamma = 1.7$. This spectrum is typical for AGN and LMXBs \citep[e.g.][]{2000MNRAS.316..234R,2003ApJ...587..356I,2005A&A...432...15P}.   We co-added the individual images and exposure maps to generate a large mosaic of the Coma cluster. The exposure corrected broad-band image is presented in  Figure~\ref{fig:5}. 

To identify point sources we ran the CIAO \textsc{wavdetect} tool on the merged images. To this end, we also generated maps of the point spread function for each observation with the \textsc{mkpsfmap} tool. Similarly to the images, we also merged the individual point spread function maps. We searched for point sources on multiple scales by running \textsc{wavdetect} with the wavelet scales of the square root two series from $\sqrt2$ to $32$. The point source lists contain 211 and 395 sources in the soft and broad band, respectively. To derive the flux of the detected sources, we used the power-law spectrum with a slope of $\Gamma = 1.7$ and the line-of-sight column density of the Coma cluster. To account for the background emission, we used local regions around each point source. The applied background regions were elliptic annuli, whose inner and outer radii were two and three times the radius of the source region. This approach allows us to simultaneously account for both the instrumental and sky background components. This latter background component includes the unresolved emission from cosmic X-ray background sources, the local foreground emission, and most notably the large-scale emission from the Coma cluster, which exhibits significant spatial variations. 
 
The central regions of the Coma cluster are dominated by an extremely bright core \citep[e.g.][]{1992A&A...259L..31B,2012MNRAS.421.1123C}. In these regions, the X-ray surface brightness of the intracluster medium largely exceeds the emission level expected from various X-ray emitting components of UDGs. Therefore, we excluded the central $4'$ of the Coma cluster. This region approximately represents the cluster core with a flat surface brightness profile. Beyond this region the surface brightness of the  intracluster medium rapidly drops and plays a less significant role. We note that this region includes only 7 UDGs, hence excluding the inner $4\arcmin$ does not affect our analysis in any significant way.  

\begin{figure}[t!]
	\begin{center}
		\leavevmode
		\epsfxsize=.48\textwidth \epsfbox{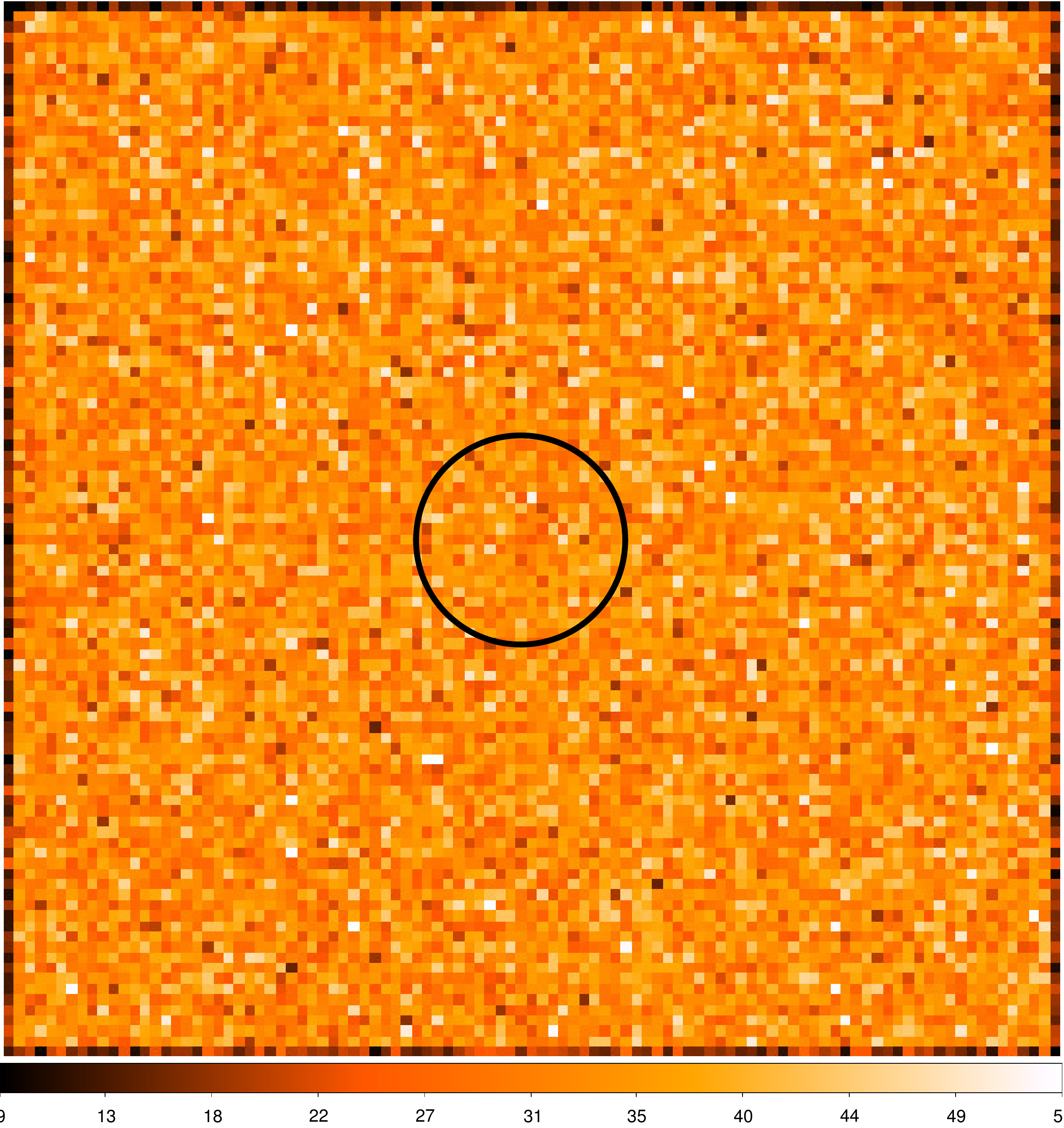}
		\caption{ Stacked soft-band image of the analyzed Coma cluster UDGs. The image covers a region of $49.2\arcsec \times 49.2\arcsec$. The circle with a radius of $5''$ marks the location of the co-added UDGs. We do not detect a statistically significant signal from the gaseous halo of the UDGs, which suggests that most UDGs reside in dwarf-sized dark matter halos.}
		\label{fig:stacked-soft-img}
	\end{center}
\end{figure}

\section{Constraining the formation scenarios of UDGs}

In this section we probe whether the observed X-ray luminosities of Coma cluster UDGs are consistent with a scenario, in which UDGs reside in Milky Way-type dark matter halos. If they reside in massive dark matter halos, UDGs should host a substantial amount of hot ionized gas and a copious population of LMXBs associated with their GC population. The total predicted X-ray emission is the combination of these (and possibly other) components   \citep[e.g.][]{2004MNRAS.349..146G,2008MNRAS.388...56B,2011MNRAS.418.1901B, 2011ApJ...729...12B}. In this work, we employ a conservative approach and assume that in the soft and broad bands the main contribution originates from the hot gas and LMXBs, respectively.

\subsection{X-ray emission from hot gaseous halos}
\label{sec:hotgas}

To test the formation scenarios of UDGs, we first probe their hot gas content. The two different evolutionary scenarios imply drastically different halo masses (Section \ref{sec:intro}). Since the X-ray gas content and hence the X-ray luminosity of galaxies is proportional to the total gravitating mass of galaxies \citep{2013ApJ...776..116K,2018ApJ...857...32B}, measuring the X-ray luminosity of UDGs directly constrains the dark matter halo mass. Here, we carry out a similar analysis to that of \citet{2019ApJ...879L..12K}, where we explored the X-ray gas content of isolated UDGs.

If UDGs are the descendants of massive galaxies, they are expected to live in Milky Way-type dark matter halos with $\sim 10^{12} \ \rm{M_{\odot}}$. This picture is supported by observations of several UDGs, such as DFX1 or Dragonfly~44 in the Coma cluster, which suggest that these galaxies reside in dark matter halos with masses of $M_{\rm vir} = (5-8)\times10^{11} \ \rm{M_{\odot}}$ \citep{2016ApJ...828L...6V,2017ApJ...844L..11V}.  To estimate the expected gaseous X-ray luminosity of an UDG with a massive dark matter halo, we assume a virial mass of $M_{\rm vir} = 8\times10^{11} \ \rm{M_{\odot}}$, which corresponds to a total mass of  $M_{\rm tot} = 1.8\times10^{11} \ \rm{M_{\odot}}$ within $5r_{\rm eff}$.  Based on the $L_{0.3-8\rm{keV}} - M_{\rm tot}$  scaling relation, the expected X-ray luminosity of the gaseous halo is $L_{0.3-8\rm{keV}} \approx 1.8\times 10^{39} \ \rm{erg \ s^{-1}} $. To convert this luminosity to the $0.5-1.2$~keV band, we  assumed a gas temperature of $kT = 0.2$~keV and a metallicity of $Z=0.2 \ \rm{Z_{\odot}}$, which results in $L_{0.5-1.2\rm{keV}} \approx 1.1\times 10^{39} \ \rm{erg \ s^{-1}} $. While the \textit{Chandra} data of most individual galaxies is not sufficiently deep to detect this relatively low luminosity, a statistically significant signal may be obtained when the full sample of galaxies is stacked. Indeed, by stacking the X-ray photons associated with individual galaxies, we increase the signal-to-noise ratios, which allows us to probe the X-ray emission of the average UDG with better sensitivity.

We carry out the stacking analysis by utilizing the soft band images and exposure maps, as we expect that the hot gas associated with the UDGs is relatively cool ($kT\sim0.2$~keV), i.e.\ most of the emission falls in the soft band. To stack the individual galaxies, we crop a $49.2\arcsec \times 49.2\arcsec$  (or $\sim$\,$23.4\ \rm{kpc}\times 23.4 \ \rm{kpc}$  at the distance of the Coma cluster) region of the image and exposure map around each UDG. As the next step, we co-added the cropped images and exposure maps that were centered on the UDG coordinates as defined by the Subaru survey.  The co-added soft-band image is presented in Figure~\ref{fig:stacked-soft-img}.

\begin{figure}[!t]
	\begin{center}
		\leavevmode
		\epsfxsize=.48\textwidth \epsfbox{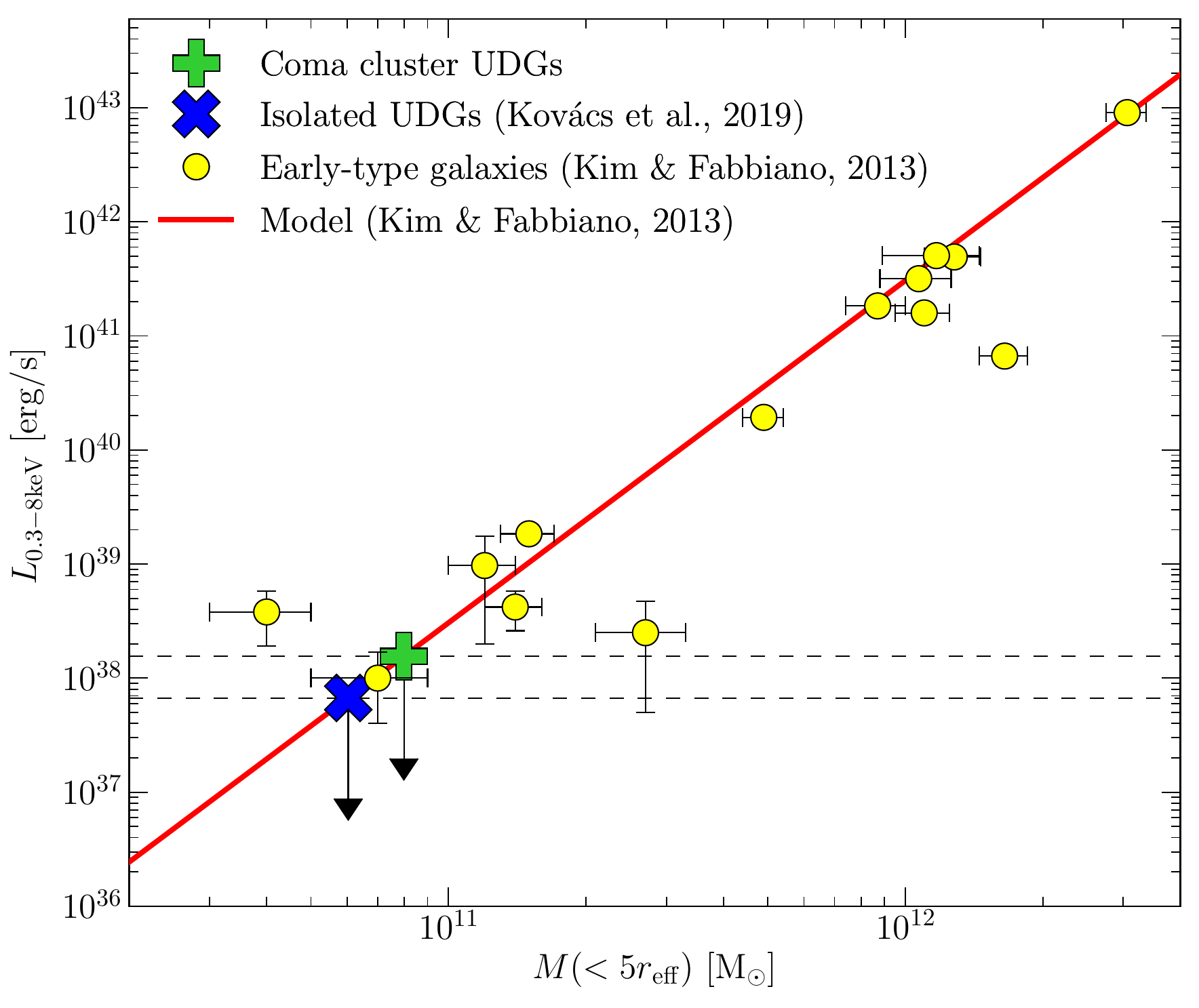}
		\vspace{-.3cm}
		\caption{Correlation between the $0.3 - 8$~keV band X-ray luminosity of hot gas and the total gravitating mass within $5r_{\rm eff} $ (red line). The upper limit on the X-ray luminosity of Coma cluster UDGs suggests a total mass of $M(<5r_{\rm{eff}}) < 8.0\times10^{10} \ \rm{M_{\odot}} $.  This upper limit is similar, albeit somewhat higher than that obtained for field UDG candidates  \citep{2019ApJ...879L..12K}.  The scaling relation predicts that galaxies with $M_{\rm vir} = 8 \times 10^{11} \ \rm{M_{\odot}} $ (or $M(<5r_{\rm{eff}}) < 1.8\times10^{11} \ \rm{M_{\odot}} $) have an X-ray luminosity of $L_{\rm 0.3-8keV} = 1.8\times10^{39} \ \rm{erg \ s^{-1}}$, which exceeds the upper limit obtained for Coma cluster UDGs by a factor of about 12.}
		\label{fig:hotgas}
	\end{center}
\end{figure}

To analyze the stacked images, we used a circular aperture with a $5\arcsec$ radius. The background region was a circular annulus with radii of $10\arcsec-15\arcsec$. After accounting for the background emission, we did not detect a statistically significant signal from the stacked sample of UDGs. To place an upper limit on the X-ray luminosity of the hot gas, we took into account the source and background counts and the stacked exposure maps. We obtain a $2\sigma$ upper limit on the flux $F_{\rm 0.5-1.2kev}<7.2\times10^{-17} \ \rm{erg \ s^{-1} \ cm^{-2} }$, which -- at the distance of Coma -- corresponds to a luminosity upper limit of $L_{\rm 0.5-1.2keV}<9.1\times10^{37} \ \rm{erg \ s^{-1} }$. This value is $\sim12$ times lower than that expected from a galaxy with a Milky Way-type dark matter halo (Figure \ref{fig:hotgas}). This suggests that most UDGs in the Coma cluster reside in dwarf-size dark matter halos. We note that this conclusion is in good agreement with our previous study for field UDGs.  

\begin{figure*}[t!]
	\begin{center}
		\leavevmode
		\epsfxsize=.49\textwidth \epsfbox{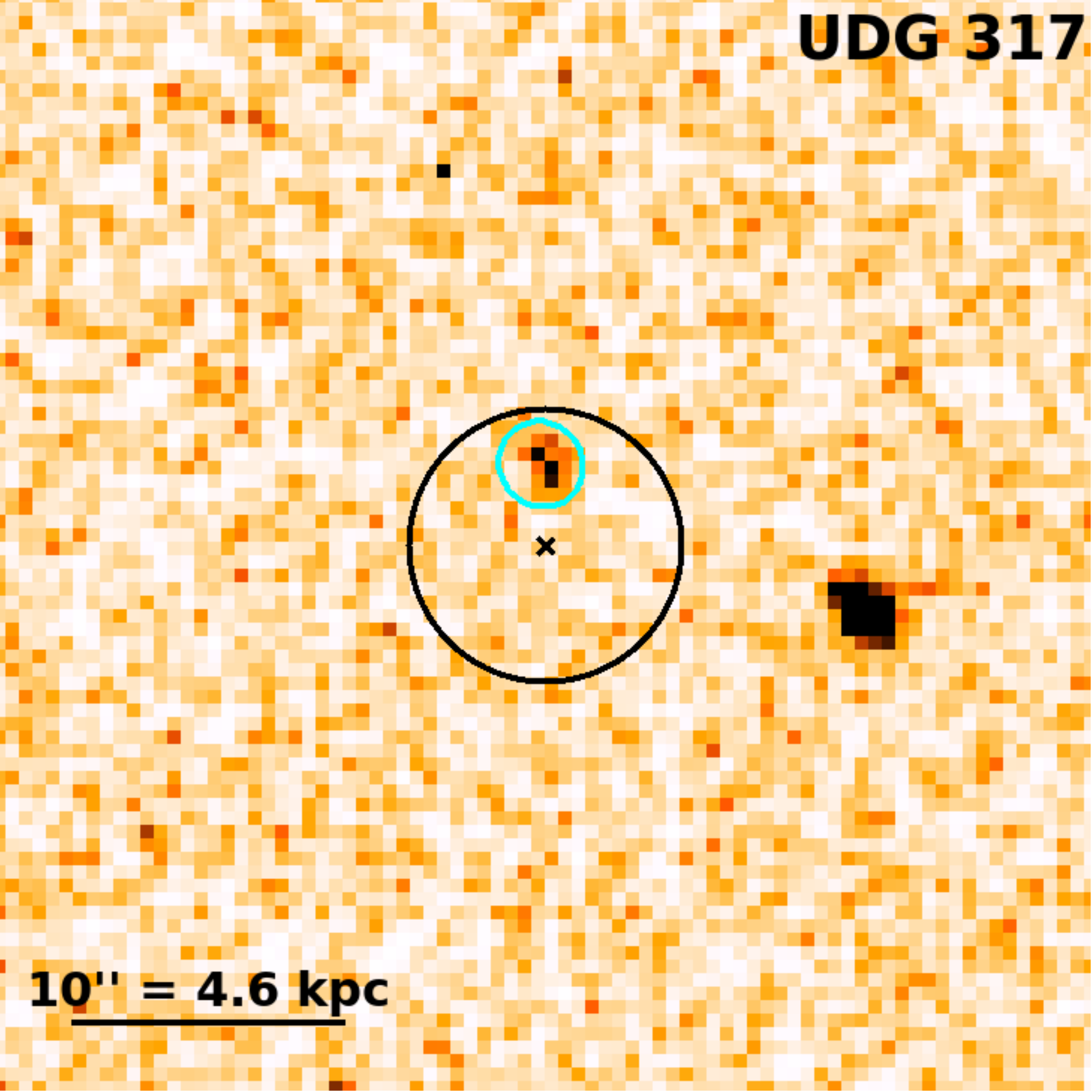}
		\epsfxsize=.49\textwidth \epsfbox{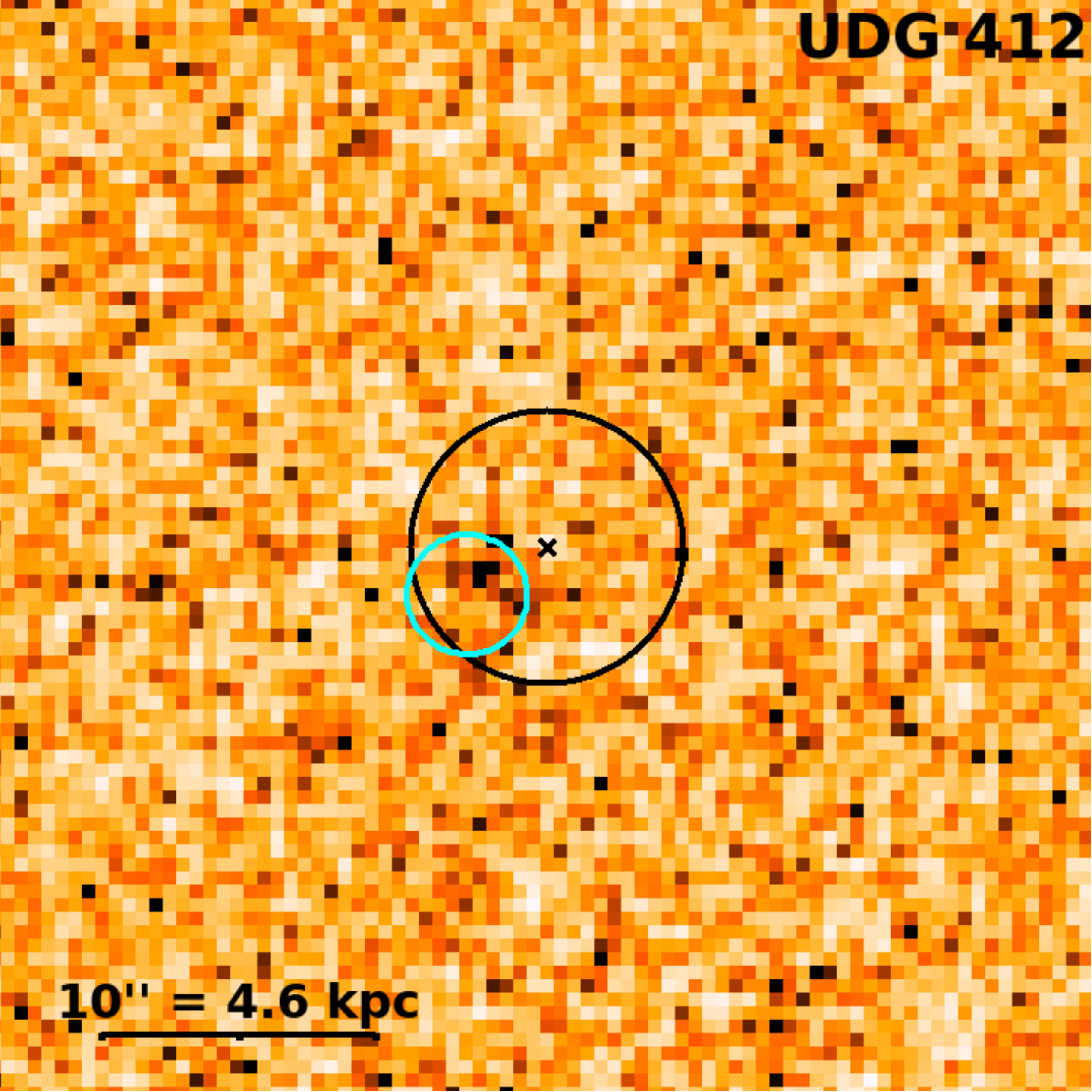}
		\vspace{0cm}
		\caption{$0.5-7$~keV band exposure corrected \textit{Chandra} images of the regions around UDG\,317 and UDG\,412 in the Coma cluster. The large circular region has a radius of $5\arcsec$ and is centered on the UDGs (shown with the cross). The small cyan ellipses show the position of the X-ray sources. The offset between the center of the UDGs and the X-ray sources is $3.0\arcsec$ and $3.2\arcsec$ for UDG\,317 and UDG\,412, respectively. Assuming that the X-ray sources reside in the Coma cluster, their broad band luminosities are $(1.3\pm0.1) \times10^{39} \ \rm{erg \ s^{-1}}$ and  $ (2.5\pm0.1) \times10^{39} \ \rm{erg \ s^{-1}}$, respectively.}
		\label{fig:6}
	\end{center}
\end{figure*}

\subsection{X-ray emission from GC-LMXBs}
\label{sec:lmxbs}

A major surprise about UDGs was the detection of a large globular cluster (GC) population. For example, Dragonfly~44 and DFX1  host $74\pm18$ and $62\pm17$ GCs, respectively \citep{2017ApJ...844L..11V}. This significant GC population exceeds the number of GCs typically found in dwarf galaxies, which host $\lesssim10$ GCs \citep{2011EAS....48..261G}, and is comparable to massive galaxies, such as the Milky Way \citep{1996AJ....112.1487H}. The large number of GCs around UDGs supports the formation scenario, in which UDGs are descendants of massive galaxies and reside in massive dark matter halos.

X-ray studies of nearby galaxies established that the number of LMXBs per unit stellar mass is significantly higher in GCs than in the galactic field\footnote{In this context, galactic field LMXBs originate from the evolution of isolated binaries, which sources are typically located in the stellar body of the host galaxy.}, which is attributed to different LMXB formation scenarios in the field and in GCs. Field LMXBs form through the primordial evolutionary path \citep{1998ApJ...493..351K}, while LMXBs residing in GCs form through dynamical interactions due to the dense stellar environment \citep{2006A&A...447...71V}. Therefore, the distribution of field LMXBs follows the stellar mass \citep{2004MNRAS.349..146G} and GC-LMXBs are distributed following the $\rho_{\star}^2/v$ law \citep{1975MNRAS.172P..15F}. Thanks to deep \textit{Chandra} observations, the LMXB population of nearby galaxies has been studied to a great extent, and it was established that the  X-ray luminosity function of field and GC-LMXBs are different \citep[e.g.][]{2011A&A...533A..33Z,2016ApJ...818...33P}. Specifically, the luminosity function of GC-LMXBs is flatter at the faint end ($<10^{37} \ \rm{erg \ s^{-1}}$) and the fraction of faint sources is factor of about 4 times less than that in the field population. 

Given their low stellar mass and the lack of a dense stellar environment, we do not expect a substantial population of field LMXBs in UDGs. However, if UDGs reside in massive dark matter halos and host a substantial GC population, the LMXBs residing in GCs are expected to have significant X-ray emission. The X-ray luminosity of LMXBs is in the the range of $10^{35} - 10^{39} \ \rm{erg\ s^{-1}}$ with only a small fraction of the GC-LMXBs exceeding $10^{38} \ \rm{erg \ s^{-1}}$ \citep{2011A&A...533A..33Z}. Therefore, given the presently available \textit{Chandra} data, most LMXBs remain unresolved at the distance of the Coma cluster, and their emission contributes to the overall diffuse emission. When the emission from individual unresolved LMXBs is combined, we may detect significant X-ray luminosity. By confronting the predicted luminosity from GC-LMXBS with the observed luminosity around Coma cluster UDGs, we can constrain whether UDGs host a large number of GC. This, in turn, allows us to probe whether UDGs reside in massive dark matter halos, thereby constraining their formation scenario. 

Assuming that UDGs reside in a Milky Way-type dark matter halo with $M_{\rm vir} \approx 8\times10^{11} \ \rm{M_{\odot}}$ (Section \ref{sec:hotgas}), and applying the relation between the host galaxy mass and the number of GCs \citep{2020AJ....159...56B}, UDGs should host an average of 160 GCs. Based on the average GC-LMXB luminosity function, the predicted combined $0.5-7$~keV band X-ray luminosity from GC-LMXBs is $\sim9\times10^{39} \ \rm{erg \ s^{-1}}$.

To probe the population of Coma cluster UDGs as a whole, we co-add the X-ray data associated with the individual UDGs.  The stacking procedure is identical with that outlined in Section \ref{sec:hotgas}, but the analysis is carried out for the broad band. We compute the luminosity associated with the stacked UDGs in a circular aperture with a radius of  $5\arcsec$. To account for the background emission, we utilize a circular annulus with radii of $10\arcsec-15\arcsec$.  Given that the typical extent of GCs from the center of galaxies is $\sim$\,$2.2 r_{\rm eff}$ \citep{2017ApJ...844L..11V}, and the average effective radius of the analyzed UDGs ($r_{\rm eff} = 2.5 \arcsec$), we expect that most GCs will be confined within this  aperture. We did not detect a statistically significant signal within the source aperture. In the absence of a detection, we compute a $2\sigma$ upper limit. Assuming a typical LMXB power-law spectrum, we obtained the flux upper limit of $F_{\rm 0.5-7keV}<8.7\times10^{-17}\, \rm erg \, s^{-1} \, cm^{-2}$, which corresponds to a luminosity upper limit of $L_{\rm 0.5-7keV}<1.1\times10^{38}\, \rm erg \, s^{-1}$. We note that this is a conservative upper limit on the luminosity from GC-LMXBs since other sources may also  contribute to the unresolved X-ray signal with the most notable emission expected from the population of field LMXBs and HMXBs. However, as discussed in \citet{2019ApJ...879L..12K}, the X-ray emission from these sources is on the order of $10^{35} - 10^{36} \ \rm{erg \ s^{-1}}$, which is several orders of magnitude lower than the observed upper limit, implying that the emission from these sources does not affect our conclusions.

The upper limit on the luminosity is a factor of $\gtrsim 80$ times lower than the luminosity predicted from a large population of GC-LMXBs.  This suggests that most UDGs do not host a significant population of GCs. Taking the upper limit on the X-ray luminosity at the face value, we estimate that UDGs host a small ($\sim1.6$) GC population, which would imply a halo mass of $M_{\rm vir}<8.3\times10^{10} \ \rm{M_{\odot}}$, consistent with dwarf galaxy halos \citep{2017MNRAS.467.2019R}.

\section{AGN occupation fraction of UDGs}
\label{sec:udgoccupation}

\begin{figure*}[ht!]
	\begin{center}
		\leavevmode
		\epsfxsize=.99\textwidth \epsfbox{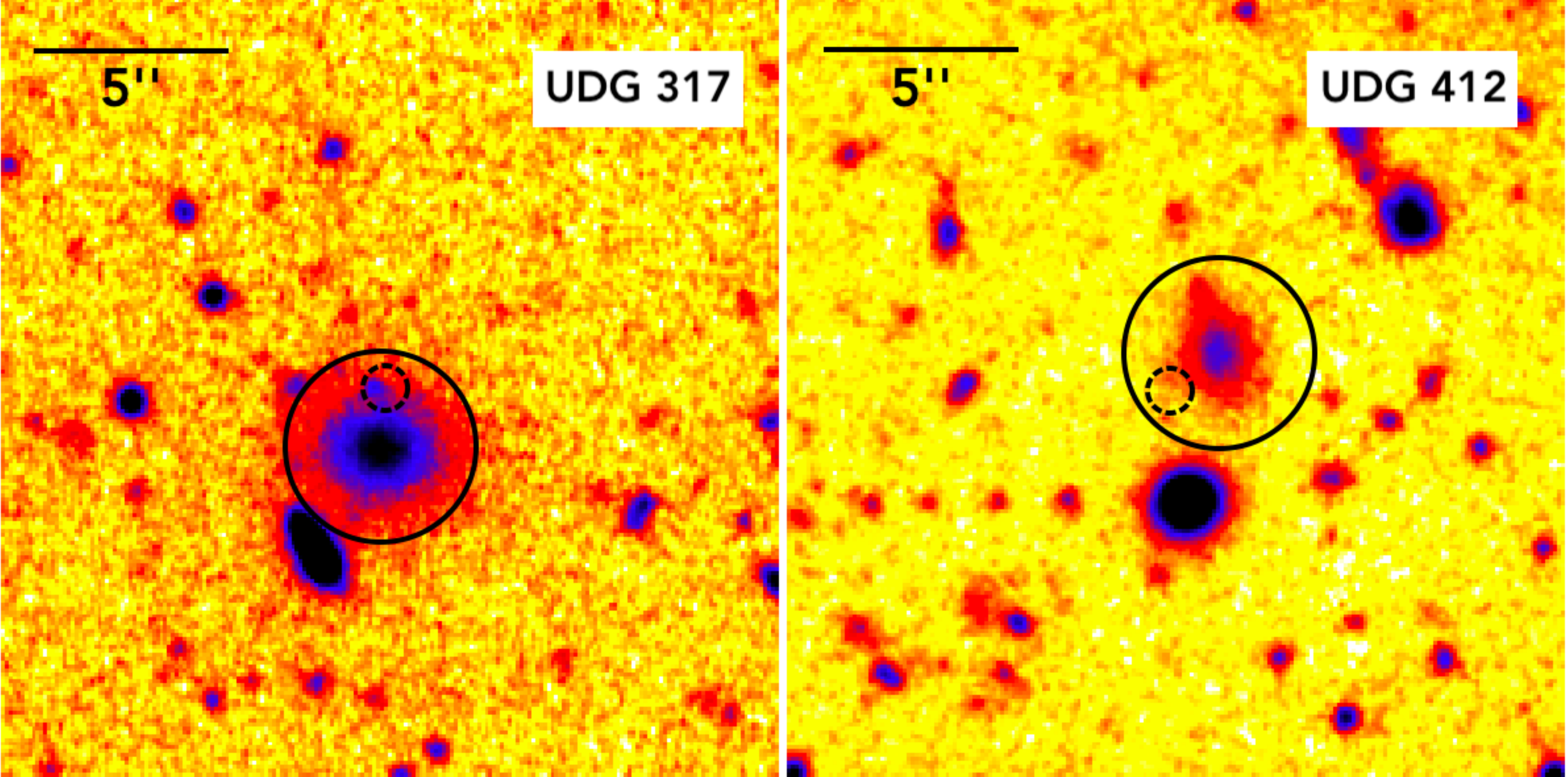}
		\vspace{0cm}
		\caption{\textit{Subaru} $B$-band images taken from the region around UDG\,317 and UDG\,412 (circles with solid lines) with a matching X-ray point source (circles with dashed lines). The X-ray source associated with UDG\,317 has an optical counterpart (\textit{left panel}), however, the X-ray source associated with UDG\,412 (\textit{right panel}) does not reveal an optical source. Since accurate redshift measurements are not available for the galaxies and for the X-ray/optical sources, these sources could either be off-center AGN in the UDGs or background AGN behind the Coma cluster.}
		\label{fig:4}
	\end{center}
\end{figure*}

\subsection{Searching for AGN in UDGs}

To search for AGN in UDGs, we cross-correlated the coordinates of the detected X-ray point sources in the Coma cluster with the position of the UDGs in both the soft and broad bands. When matching the X-ray sources with the center of galaxies, a search radius needs to be defined. The search radius depends on the positional accuracy of the \textit{Chandra} point source detection, which is determined by the number of source and background counts, and the off-axis distance. In-depth studies demonstrate that the typical position offset between the center of SDSS galaxies and X-ray sources is $\sim0.7\arcsec\pm0.4\arcsec$. Moreover, for $\gtrsim97\%$ of the sources the offset is $<2.5\arcsec$ even for those with a large off-axis distance \citep{2007ApJS..169..401K,2012ApJS..200...17T}. Based on these, we conservatively use a search radius of $2.5\arcsec$. After cross-correlating the X-ray source positions with the central coordinates of UDGs, we did not find any matches within $2.5\arcsec$. 

Because UDGs have relatively flat surface brightness profiles \citep[e.g.][]{2016ApJ...828L...6V}, and simulations and observations suggest that a notable fraction of dwarf galaxies have BHs that are wandering within a few kpc of the galaxy center \citep{2019MNRAS.482.2913B,2020ApJ...888...36R}, we considered whether UDGs may host off-center BHs. Therefore, we increased the search radius to $5\arcsec$. Given the typical effective radius ($r_{\rm eff} =2.5\arcsec$) of UDGs in our sample, this search radius broadly corresponds to $2r_{\rm eff}$ at the distance of Coma cluster. 

Using the large, $5\arcsec$, search radius, we identified two galaxies, UDG\,317 ($r_{\mathrm{eff}}=2.7\arcsec$) and UDG\,412 ($r_{\mathrm{eff}}=1.8\arcsec$), which match with a detected X-ray source. The \textit{Chandra} images of these two galaxies and the associated X-ray point sources are shown in Figure~\ref{fig:6}.  We note that the relatively extended nature of the point sources is due to the $\sim4\arcsec$ size of the weighted point spread function at the location of both X-ray sources. The offset between the galaxy centroid and the X-ray source positions is $3.0\arcsec$ and $3.2\arcsec$ for UDG\,317 and UDG\,412, respectively. At the distance of the Coma cluster, these offsets correspond to $\sim1.4$ kpc and $\sim1.5$ kpc, respectively. We note that the detection of offset AGN may not be unusual in dwarf galaxies. Indeed, cosmological simulations  \citep{2019MNRAS.482.2913B} of dwarf galaxies  suggest that  $\sim50\%$ of dwarfs may host their ``central" BHs off-center, of which one-third lies at $\gtrsim1$\,kpc off-center distance, mostly concentrated between $1-1.5$\,kpc.

\subsection{The X-ray sources in UDG\,317 and UDG\,412}

Assuming that the X-ray point sources are in the Coma cluster, the observed net count rates of $( 3.7\pm0.3) \times10^{-5} \ \rm{s^{-1}}$ and $( 5.5\pm 0.3) \times10^{-5} \ \rm{s^{-1}}$ correspond to luminosities of $L_{\rm 0.5-7keV} =(1.3\pm0.1) \times10^{39} \ \rm{erg \ s^{-1}}$ and  $L_{\rm 0.5-7keV} = (2.5\pm0.1) \times10^{39} \ \rm{erg \ s^{-1}}$ for UDG\,317 and UDG\,412, respectively. If these UDGs reside in the Coma cluster and the X-ray sources are within the galaxies, they are most likely off-center low-luminosity AGN. Indeed, the luminosity of both X-ray sources exceeds $10^{39} \ \rm{erg \ s^{-1}}$, which is above the cut-off luminosity of low-mass X-ray binaries  \citep{2004MNRAS.349..146G}. While high-mass X-ray binaries or ultra-luminous X-ray (ULXs) sources may reach luminosities in excess of $10^{39} \ \rm{erg \ s^{-1}}$, these types of sources are associated with active star formation \citep{2003MNRAS.339..793G,2004MNRAS.347L..18K,2009ApJ...703..159S,2012MNRAS.419.2095M}. However, UDG\,317 and UDG\,412 have color indeces of $B-R=0.91$ and $B-R=0.92$, which suggests that these are passive galaxies without significant star formation \citep{2015ApJ...807L...2K}. Therefore, it is unlikely that any of these sources are HMXBs or ULXs in these galaxies. As a caveat, we note that the galaxies or associated X-ray sources could be projected onto the Coma cluster. For example, if the galaxies and their point sources are foreground galaxies that reside at $D\lesssim65$~Mpc, the luminosity of the off-nuclear X-ray sources remains below $10^{39} \ \rm{erg \ s^{-1}}$. This luminosity could be explained by LMXBs. 

Due to the low number of net counts ($\lesssim 100$) associated with the X-ray sources, it is not possible to construct their X-ray energy spectrum. Therefore, we computed simple hardness ratios to infer the nature of these sources. Since we detect $<10$ net counts in the soft band, we rely on counts measured in the hard ($2-7$\,keV) and medium ($1.2-2$\,keV) band. We define the hardness ratio as  $\rm{HR}  = (F_{\rm hard}-F_{\rm medium}) / (F_{\rm hard}+F_{\rm medium})$, where $F$ is the exposure corrected number of counts in the given energy band. For the X-ray sources in UDG\,317 and UDG\,412, we obtained hardness ratios of $\rm{HR} = 0.23 $ and $\rm{HR} = 0.20 $, respectively. The typical spectrum of AGN can be described with a power-law model with slopes of $\Gamma = 1.5-2$, which result in hardness ratios of $\rm{HR}_{\rm model}  = 0.04-0.23 $, where the lower value corresponds to the slope of $\Gamma = 2$. Thus, the observed hardness ratios are consistent with a power-law model with a slope of $\Gamma \sim 1.5-1.7 $, suggesting that the X-ray sources in UDG\,312 and UDG\,412 could originate from AGN. 

UDG\,317 and UDG\,412 are covered by 24 and 11 \textit{Chandra} pointings, which allows us to probe whether the X-ray sources associated with these galaxies originate from multiple observations. We find that both X-ray sources are present in multiple X-ray images, albeit they are not detected individually in each observation. Thus, these sources are not only detected in a single observational epoch due to their stochastic brightening, such as an AGN outburst. However, a detailed  investigation of the temporal variations of the X-ray sources is not possible due to the low number of counts and the consequently large  statistical uncertainties. 

To further investigate the X-ray sources in UDG\,317 and UDG\,412, we inspect the Coma cluster images taken by the \textit{Subaru} \citep{2016ApJS..225...11Y}; these images are presented in Figure \ref{fig:4}. Interestingly, we identify an optical source that is coincident with the X-ray source identified in UDG\,317. The presence of the optical counterpart suggests that this source is an AGN. Indeed, if the X-ray source was an X-ray binary or a ULX, the donor star would be too faint to be detected at the distance of the Coma cluster. However, it is still possible that the AGN is not residing in the galaxy, but it is a foreground or, more likely, a background object. To conclusively determine whether this source is truly an off-center AGN associated with UDG\,317, the redshift of the UDG and the optical source should be measured. We do not detect an optical counterpart of the X-ray source in UDG\,412. The non-detection of an optical source does not constrain the nature of the X-ray source. Specifically, it could be an AGN residing in the UDG, a high-redshift background AGN, or a luminous (foreground) X-ray binary. Thus, based on the present data the true nature of these X-ray sources cannot be conclusively determined.  Therefore, we plan to carry out an optical follow-up of UDG\,317 and UDG\,412 along with the associated sources, which will be subject to a future investigation. 
 
 \subsection{AGN occupation fraction of the Coma cluster UDGs}

In this work, we identified luminous X-ray sources associated with UDG\,317 and UDG\,412. However, given the present data, the nature of these X-ray sources cannot be unequivocally constrained. In addition, we  did not detect an X-ray source associated with the other 402 Coma Cluster UDGs in the \textit{Chandra} footprint. Therefore, we place an upper limit on the AGN occupation fraction of UDGs.  Based on the two  \textit{potential} off-center AGN and the non-detections, the upper limit on the AGN occupation fraction of Coma cluster UDGs is $f_{\rm occ} <0.5\%$.

\begin{figure}[t]
	\begin{center}
		\leavevmode
		\epsfxsize=.5\textwidth \epsfbox{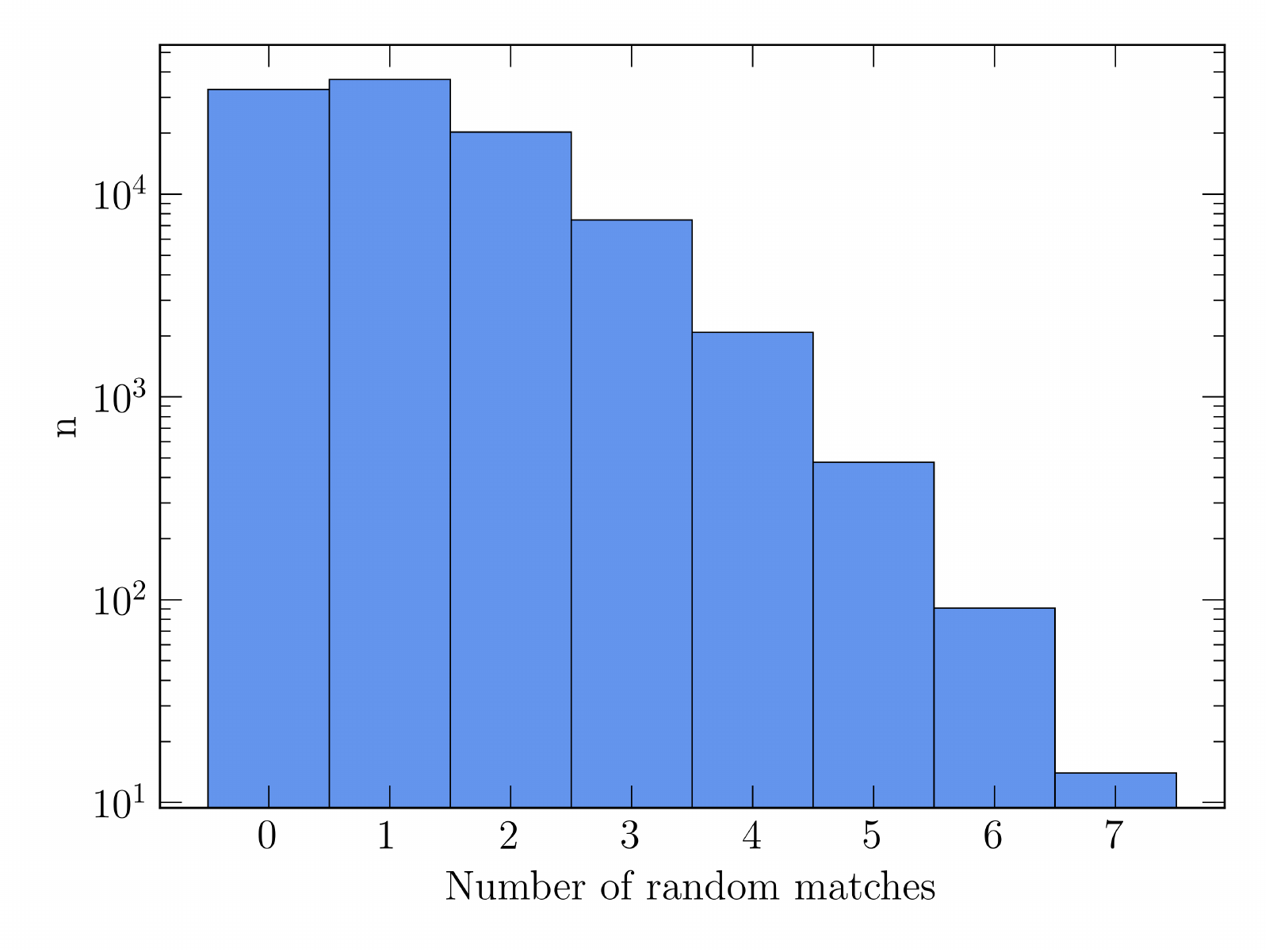}
		\caption{Results of Monte Carlo simulations showing the number of random X-ray source-UDG matches in the field of Coma cluster for $10^5$ randomly generated UDG coordinate sets. We conclude that $69\%$ of simulations results in $0-1$ matches out of 404 randomly selected coordinates, which suggests that one of the two UDG-X-ray source pairs detected is likely the result of spatial coincidence with a background source.}
		\label{fig:8}
	\end{center}
\end{figure}

\subsection{Monte Carlo simulations}
\label{sec:mc}

Due to the large number of UDGs and the abundance of X-ray point sources in the Coma cluster, it is possible that some of the X-ray point sources associated with UDGs are due to spatial coincidence. In this scenario, the X-ray sources are not AGN in the galaxy but are background AGN.

To assess the likelihood of random matches, we carried out Monte Carlo simulations. To this end, we randomly generated $404$ coordinates within the \textit{Chandra} footprint of the  Coma cluster excluding the central $4\arcmin$ region. Hence the number of random coordinates is the same as the number of UDGs used in our work. Then, using these random coordinates, we searched for matching X-ray sources within $5\arcsec$ between the position of the detected X-ray sources and the random coordinates. To build a statistically meaningful sample, we repeated this experiment $10^5$  times, each time for a different set of random coordinates. We recorded the number of matches for each simulation and present the results on a histogram in Figure \ref{fig:8}. The number of random matches  ranges between $0-7$ with a mean of $1.11$ and a median of $1.0$, which corresponds to a mean chance occupation fraction of $f_{\rm occ}^{\rm MC}=0.27\%$. We find that $\sim 69\%$ of the simulations have $0$ or $1$ random match and $\sim20\%$ have two spatially coincident matches. Based on these, it is likely that at least one, but possibly both, UDG-X-ray source pairs are the results of spatial coincidence.

\section{Discussion}
\label{sec:discussion}

In this work, we probed  the formation scenarios of UDGs by constraining their typical dark matter halo mass  using multiple approaches. First, we probed the hot X-ray gas content of UDGs. We did not detect a luminous gaseous halo around individual galaxies and in the stacked sample, which demonstrates the absence of hot X-ray gas. This result resonates well with our earlier study of isolated UDGs \citep{2019ApJ...879L..12K}, and suggest that most UDGs reside in dwarf-sized dark matter halos in isolation and in rich environments. 

If UDGs reside in massive dark matter halos, they should also host a significant GC population and hence a substantial number of GC-LMXBs. We constrained the X-ray emission originating from LMXBs residing in GCs. As we did not detect a statistically significant signal from GC-LMXBs, we placed an upper limit on the X-ray luminosity of LMXBs, which suggested that the typical UDGs host $\sim1.6$ GCs.  We note that the upper limit on the number of GCs was derived based on the best-fit scaling relation between the total mass of galaxies and the number of GCs. Although this scaling relation is very tight, at low virial masses ($M_{\rm vir} \lesssim 10^{11} \ \rm{M_{\odot}} $) it exhibits somewhat larger scatter \citep{2020AJ....159...56B}. However, this scatter does not influence our conclusions in any significant way, since we stacked a large sample of UDGs, which approach probes the average galaxy properties and guards against outliers. The low number of inferred GCs for UDGs is comparable with that obtained for dwarf galaxies and is about $1.5-2$ orders of magnitude lower than that observed for massive galaxies.  We also emphasize that the applied scaling relation between the total mass of galaxies and the number of GCs is empirical and is not based on a volume-limited galaxy sample, however, this work represents the largest published survey. We refer the reader to \citet{2020AJ....159...56B} for a discussion about details on potential biases. Overall, our results suggests that most UDGs live in dwarf-sized dark matter halos and are genuine dwarf galaxies. However, as a caveat, we mention that the formation efficiency of LMXBs depends on the metallicity of GCs \citep{2003ApJ...589L..81K}. Specifically, metal-rich red GCs are at least $3$ times more efficient in forming LMXBs than metal-poor blue GCs. It is known that the stellar body of galaxies host red GCs, while the dark matter halo hosts blue GCs. Given that in UDGs we mostly expect blue GCs \citep{2016ApJ...830...23B,2017ApJ...844L..11V}, which are associated with the dark matter halo, we may expect a significantly lower LMXB formation efficiency. However, even if the formation efficiency is a factor of 3 lower, the difference between the observed and predicted X-ray luminosities is still a factor of $10-30$. Thus, it is unlikely that most UDGs host a significant GC population.

In summary, our analysis of the Coma cluster UDGs strengthens the formation scenario, in which UDGs are the puffed-up version of dwarf galaxies. Based on the present analysis and the results presented in  \citet{2019ApJ...879L..12K}, we suggest that most UDGs are genuine dwarf galaxies and originate from similar evolutionary scenarios in both isolated and rich environments. However, it cannot be excluded that a small subset of UDGs, such as Dragonfly 44, resides in extended dark matter halos. Deep X-ray observations with present-day X-ray telescopes can provide an independent method to measure the dark matter halo mass of individual galaxies. In addition, future X-ray observatories with large collecting areas and high spatial resolution, such as the proposed Lynx observatory, will be excellent tools to understand UDGs. Such future studies will allow the study of both larger samples of UDGs over greater distances and will also enable us to accurately distinguish the X-ray emitting components within these systems.

In this work, we also placed an upper limit on the AGN occupation fraction of UDGs. It is interesting to compare these results with the AGN occupation fraction obtained for other galaxies. Based on SDSS observations, \citet{2013ApJ...775..116R} studied a sample of about 25000 emission-line galaxies with $z<0.055$ (or $D<250$~Mpc) with stellar masses of  $ 10^{8.5} \lesssim M_{\star} \lesssim 10^{9.5} \ \rm{M_{\sun}}$. While this sample represents the population of dwarf galaxies, these galaxies are more massive than the UDGs in our study. They identified 151 galaxies that potentially host an AGN, implying an AGN occupation fraction of $\sim0.5\%$. As a caveat, we note that the sample of  \citet{2013ApJ...775..116R}  consists of optically detected AGN, therefore they can only identify relatively luminous AGN, with high accretion rates. 

In the framework of the  AMUSE (AGN Multiwavelength Survey of Early-type galaxies) survey,  \citet{2015ApJ...799...98M} probed the AGN occupation fraction of about 200 optically-selected early-type galaxies. The stellar mass of the galaxies was in the range of $ 10^{7.6} \lesssim M_{\star} \lesssim 10^{12} \ \rm{M_{\sun}}$ and they revealed that the AGN occupation fraction in galaxies increases with stellar mass. For their lowest mass bin, $ 10^{7.6} \lesssim M_{\star} \lesssim 10^{9.5} \ \rm{M_{\sun}}$, they obtained an AGN occupation fraction of $f_{\rm occ} \sim 5.6\%$. While this value is about an order of magnitude higher than that obtained in our study, galaxies studied in the  AMUSE survey reside at a distance of $15-27$~Mpc, which, in turn, allows the detection of fainter X-ray sources. Specifically, our typical source detection sensitivity is few times $10^{39} \ \rm{erg \ s^{-1}}$, whereas the faintest X-ray sources in the AMUSE survey have luminosities of  few times $10^{38} \ \rm{erg \ s^{-1}}$. Therefore, the higher AGN occupation fraction obtained for dwarf galaxies may be -- at least in part -- due to the more sensitive nature of the AMUSE survey. The difference in the AGN occupation fraction is even more striking between UDGs and massive galaxies. Galaxies in the AMUSE survey with  $M_{\star} > 10^{10} \ \rm{M_{\sun}}$ exhibit an AGN occupation fraction of $f_{\rm occ} \gtrsim 70\%$, which more than two orders of magnitude higher than that obtained for UDGs in the Coma cluster. Clearly, the higher source detection sensitivity in the Coma cluster cannot account for this difference. Thus, we conclude that the AGN occupation fraction of UDGs is similar or even lower than that obtained for dwarf galaxies. 

As we discussed in Section \ref{sec:udgoccupation}, we computed the upper limit on the AGN occupation fraction based on two X-ray point source--UDG pairs. However, in the absence of precise spectroscopic redshift measurements, the distances of the galaxies and the X-ray sources are not known. Therefore, it is feasible that the matches are due to projection effects as suggested by our Monte Carlo simulations. Therefore, the X-ray point sources may not be associated with the UDGs, but could be background AGN at higher redshift. Finally, it is also possible that the UDGs and the X-ray sources are only projected to the Coma cluster, but are in the foreground. In this scenario, the intrinsic X-ray luminosity of the sources may be significantly lower, and hence the sources could be X-ray binaries. To clarify these issues, we plan to carry out spectroscopic measurements, which will establish if the UDGs reside in the Coma cluster and whether the X-ray source associated with UDG\,317 is at the same distance as the galaxy itself. However, this analysis is beyond the scope of this paper and will be the subject of a future study.

\section{Conclusions}
\label{sec:conclusions}

We performed an X-ray analysis on 404 Coma cluster UDGs identified by the Subaru Suprime-Cam survey. We probed the formation scenario of UDGs using two methods: measuring the X-ray luminosity expected from hot gas and from GC-LMXBs. In addition, we constrained the AGN occupation fraction of UDGs. Our results can be summarized as follows:

\begin{itemize}

\item We measured whether UDGs host a significant amount of hot X-ray emitting gas, which would be expected if they reside in a massive dark matter halo. We carried out a stacking analysis in the $0.5-1.2$~keV energy range but did not detect a statistically significant signal. We placed an upper limit of $L_{\rm soft}<9.1\times10^{37} \ \rm{erg \ s^{-1} }$ on the luminosity of hot gas. This upper limit falls a factor of $\sim12$ times below the luminosity expected from UDGs with massive dark matter halos.

\item We constrained whether UDGs host a significant population of GCs. To this end, we probed the X-ray emission originating from the unresolved population of LMXBs residing in globular clusters. We did not detect statistically significant X-ray emission from the stacked sample, and placed an upper limit of $L_{\rm broad}<1.1\times10^{38}\, \rm erg \, s^{-1}$. This limit is $\gtrsim80$ times lower than that predicted in a scenario where UDGs host a large GC population. 

\item We searched for AGN associated with the UDGs. We identified two X-ray source--UDG pairs within $5\arcsec$, which may be off-center AGN. Since we cannot confirm that the X-ray sources are associated with the UDGs, we place an upper limit of $f_{\rm occ} < 0.5\%$ on the AGN occupation fraction of UDGs in the Coma cluster. This value is comparable or even lower than that obtained for dwarf galaxies and falls short of the values observed for massive galaxies. 

\end{itemize}

In summary, we conclude that the bulk of the UDG population are genuine dwarf galaxies and are not the descendants of massive galaxies. Combining the results presented in this work with our earlier study, we suggest that most UDGs undergo similar evolutionary scenarios in isolated and rich environments. However, as a caveat, we mention that it is feasible that a small sub-sample of UDGs may originate from a different formation channel, and may live in massive dark matter halos. \\

\smallskip

\begin{small}
\noindent
\textit{Acknowledgements.}
The authors thank Jin Koda and Yagi Masafumi for providing the Subaru images of UDG\,317 and UDG\,412, and for the discussion on the optical sources.
The authors also thank Grant Tremblay, Imad Pasha, and Lamiya Mowla for discussions on the optical follow-up of the X-ray sources. The work presented in this paper is based in part on data collected at Subaru Telescope, which is operated by the National Astronomical Observatory. This research has made use of data obtained from the Chandra Data Archive and the Chandra Source Catalog, and software provided by the Chandra X-ray Center (CXC) in the application packages CIAO, ChIPS, and Sherpa. In this work, the NASA/IPAC Extragalactic Database (NED) has been used. Funding for the Sloan Digital Sky Survey IV has been provided by the Alfred P. Sloan Foundation, the U.S. Department of Energy Office of Science, and the participating Institutions. SDSS-IV acknowledges support and resources from the Center for High-Performance Computing at the University of Utah. The SDSS web site is www.sdss.org. \'A.B. acknowledges support from the Smithsonian Institution.

\end{small}

\bibliographystyle{apj}
\bibliography{bib2} 

\end{document}